\documentclass[twocolumn]{svjour3}          % twocolumn
\usepackage{graphicx}
\usepackage{amsmath}
\usepackage{enumitem,lipsum}
\begin{document}

\title{Reaction diffusion patterns in \emph{Pseudoplatystoma} fishes %\thanks{Grants or other notes
%about the article that should go on the front page should be
%placed here. General acknowledgments should be placed at the end of the article.}
}

%\titlerunning{Short form of title}        % if too long for running head

\author{Aldo Ledesma-Dur\'an         \and
         L. H\'ector Ju\'arez-Valencia  \and 
         Iv\'an Santamar\'ia-Holek %etc.
}

%\authorrunning{Short form of author list} % if too long for running head

\institute{Aldo Ledesma-Dur\'an  \at
              Universidad Aut\'onoma Metropolitana, Campus Iztapalapa, A.P. 55-534, 09340
 D.F, Me\-xi\-co. \\
              %Tel.: +123-45-678910\\
              %Fax: +123-45-678910\\
              \email{aldo\_ledesma@ciencias.unam.mx}           %  \\
%             \emph{Present address:} of F. Author  %  if needed
           \and
           L. H\'ector Ju\'arez-Valencia  \at
             Universidad Aut\'onoma Metropolitana, Campus Iztapalapa, A.P. 55-534, 09340
 D.F, Me\-xi\-co.    \and
Iv\'an Santamar\'ia-Holek \at
UMDI-Facultad de Ciencias, Universidad Nacional Aut\'onoma de M\'exico, Campus Juriquilla,
Quer\'etaro 76230, Mexico.  
}

%\date{Received: date / Accepted: date}
% The correct dates will be entered by the editor

\maketitle

\begin{abstract}
This paper studies how patterns derived from a system of reaction-diffusion equations may vary significantly depending on  boundary and initial conditions, as well as in the spatial dependence of the coefficients involved. From an extensive numerical study of the BVAM model, we demonstrate that the geometric pattern of a reaction diffusion system is not uniquely determined by the value of the parameters in the equation. From this result, we suggest that the variability of patterns among individuals of the same specie may have its roots in this sensitivity. Furthermore, this study analyzes briefly how the inclusion of the advection and the space dependency in the parameters of the model influences the forms of a specific pattern. The results of this study are compared to the skin  patterns that appear in \emph{Pseudoplatystoma} fishes.
% \subclass{MSC code1 \and MSC code2 \and more}
\end{abstract}

%%%%%%%%%%%%%%%%%%%%%%

\section{Introduction}

Teolostei fishes exhibit a great diversity of pigmentation patterns. These patterns depend on the localization of pigmented cells, called chromatophores,  which are originated in the neural crest\cite{painter2001models}. Those pigmented cells are mainly the xanthophores (of a yellow or orange color), melanophores (black), and iridophores (grey and silver) \cite{parichy2003pigment}. The pigment cells migrate to the skin as an oscillatory perturbation until they reach their final location. Only when this migration is over, the cells acquire their characteristic pigment. Regions of skin dominated by a specific chromatophore turn into the respective color in such a way that the combination of those different regions forms a pattern \cite{Parichy2000}.

%(of a yellow or orange color), melanophores(black), and iridophores (grey and silver) 

The mechanisms that establish the spatio-temporal organization of such patterns in fishes have not been determined conclusively. Kondo and Asai  suggested that a reaction-diffusion (RD) mechanism of the Turing-type could explain this process \cite{Kondo1995}. The idea was that a RD mechanism provides a chemical pre-pattern between the chomotophores; this pattern is fixed to the skin through  some morphogen, and, finally, cellular differentiation between regions of preponderance of one or other chromatophore gives place to the pattern on the skin of the fish. The results of that work, and of subsequent similar works that follow it, were figures that reproduce in good agreement the characteristics and the evolution of several fish patterns, see Refs. \cite{meinhardt2003algorithmic,painter1999stripe,varea1997confined}.

\begin{figure}
\includegraphics[scale=0.26]{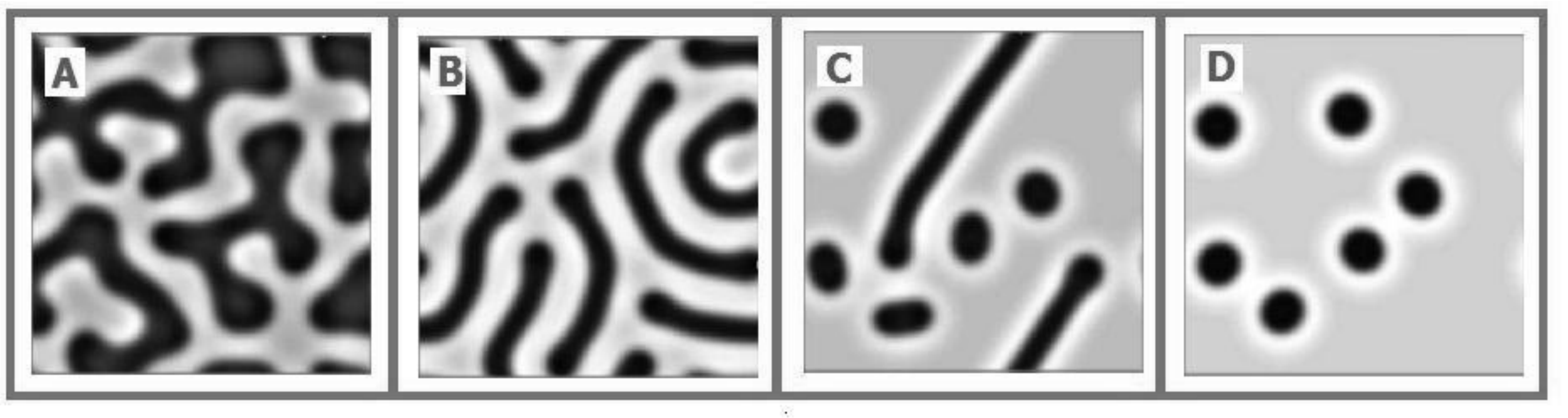}
\caption{ A change in the forms of the pattern of the BVAM model can occur by just changing  the parameter $\delta$ in equation (\ref{eq:bvam2}) : A) $\delta_1=$0.0, B) $\delta_2=$0.6 C) $\delta_3=$0.71 D) $\delta_4=$0.72. The other parameters are $\epsilon = 0.65$,  $\zeta= 0.165$, $\gamma =-2.5$, $D_0 = 0.344$ and $\eta= 0.15$. The domain is a square of size $L=80$ with zero flux boundary conditions. Taken from Ref. \cite{Barrio2009}. \label{fig:barrio}}
\end{figure}

More recently, Barrio  and co-workers suggested a possible connection between a RD mechanism, called the BVAM model, and the patterns in the surubi fishes \cite{Barrio2009}. 
They showed that changing a single parameter in the model produce several kind of patterns, such as disordered dots, combinations of lines and dots, stripes and reticulated patterns, see Figure \ref{fig:barrio}. In the same work, it was suggested the possibility of combining these basic forms in order to obtain the more elaborated patterns of the \emph{Pseudoplatystoma} fishes.

%The main result of this work was that the change of a single parameter in some specific reaction diffusion equations (called the BVAM model) can lead to the different patterns of the different species of the genus \emph{Pseudoplatystoma}.

 %Therefore, one could relate in principle this parameter with the evolutionary history of differentiation between these species, and even, in a more bold conjecture, between individuals of the same specie.

\begin{figure}
\includegraphics[scale=0.13]{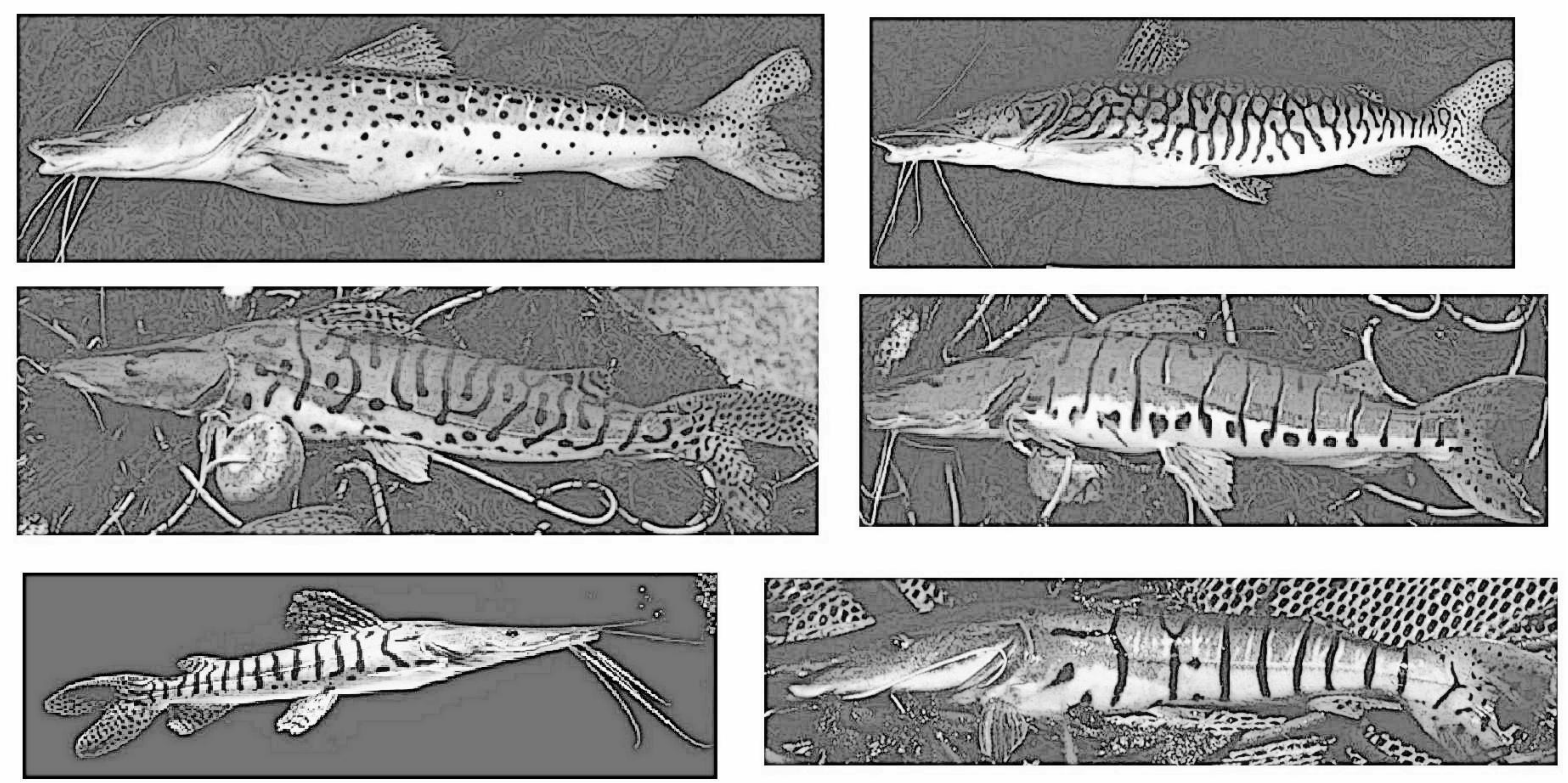}
\caption{Species of \emph{Pseudoplayistoma} Bleeker catfish. From top to bottom. Left: \emph{P. coruscans, P. fasciatum} and \emph{P.tigrinum}. Right: \emph{P. reticulatum, P. orinocoense} and \emph{P. magdaleniatum}. Taken from \cite{buitrago2007taxonomy} \label{fig:surubies}.}
\end{figure}

\emph{Pseudoplatystoma} is a group of neotropical catfishes of the family \emph{Pimelodidae} which contains three long recognized species and another possible five species recently proposed \cite{buitrago2007taxonomy}. These catfishes reach sizes that exceed 1.3 m, and live in various habitats such as large rivers, lakes and flooded neotropical forests \cite{buitrago2006anatomia}. The systematic of this small group is unknown, in part because their species have large geographic variation in morphology and coloration. However, the pigmentation of three species is  clearly different: the lateral region of the populations of \emph{P. fasciatum} has dark vertical bands; \emph{P. tigrinum} presents reticulated bands and \emph{P. coruscans} features large dots arranged in rows in the lateral region of the body \cite{buitrago2007taxonomy}, see Figure \ref{fig:surubies}.

%Its distribution includes large  riverbed of the Neotropics, Amazon, Orinoco, Paran\'a (including the Uruguay River) rivers in the region of the Guianas, San Francisco and Magdalena\cite{buitrago2006anatomia}.  They are  distinguished by the somewhat compressed head of which comes the name of the genus. Individuals of this sort are generally recognized vernacular names: "Striped Catfish", "Pintadillo" and "Caparari". 

Although the patterns obtained by Barrio and collaborators are quite similar to those present in the skin of the catfishes, there are some features that conventional reaction models, including the BVAM model, do not reproduce, for example: 1) the black and silver striped patterns on the fish have different thickness (except in the \emph{P coruscans}); 2) some fish patterns have a preferred orientation, mainly to the vertical direction;  and finally,  3) in some \emph{Pseudoplatystoma} fishes occur a variation of the geometrical forms in the vertical direction, for example, the specks occurring only at the bottom of the \emph{P. oricoense} or the reticulated on the top of \emph{P. reticulatum}; see Figure \ref{fig:surubies}. 

%This statement could be partially shown trough an extensive numerical study over a width range of the parameters synthesized in Figure \ref{fig:tapete}. 

We shall call \emph{conventional} conditions to those that arise when using 1) fixed boundaries (not growing), 2) a closed domain, 3)  constant coefficients in the reaction, and 4) random initial conditions around some fixed values (usually around the equilibrium points). Some few examples of those constraints in reaction diffusion models can be found in Refs. 
\cite{dufiet1992numerical,madzvamuse2006time,pearson1993complex,ruuth1995implicit}.

%This paper will explore how  the variation in one or more of the last three factors could contribute to the special arrangement of patterns in each kind of \emph{Pseudoplatystoma}. This means that, even though the BVAM reaction diffusion model reproduces the irregular arrangement of stripes and points these fish have, still does not explain the spatial dependence of shapes and their orientation. 

 In this work, we go beyond previous reaction-diffusion models, formulated with the aim to reproduce skin patterns observed in animals (such as the \emph{Pseudoplatystoma} catfishes), by analyzing the effect of some external factors that may influence their time evolution and final shapes. Here, by external factors we mean those which  are not included in the reaction-diffusion equations. More specifically, we will show that the morphology of the patterns depends crucially on: 1) different boundary and initial conditions; 2) the effect of advection, and 3) space dependent kinetic parameters. These factors are frequently found in real biological conditions and are very important in the spatio-temporal progress of pattern formation. Thus, even though the BVAM  model \cite{Barrio2009} reproduces the irregular arrangement of stripes and points observed on the catfishes, we will show that changing the values of the parameters of the reaction mechanism is not sufficient to explain the spatial dependence of the shapes and their orientation. On the contrary, we will show that is suffices to consider combinations of the three factors mentioned above to reproduce more readily and accurately the skin patterns observed in the \emph{Pseudoplatystoma} catfishes.

The organization of this work is the following. In Section \ref{sec:bvam}, we present the BVAM model and the basic patterns derived  from it using standard boundary and initial conditions. In Section \ref{sec:numerical}, we introduce different boundary and initial conditions for a reaction scheme with the same configuration at the parameter space, and see how the original patterns are modified. Furthermore, we will explore the effect of advection on the patterns and the spatial dependency of the kinetic parameters of the reaction in the BVAM model. In Section \ref{sec:comparison} we will compare the patterns obtained in the previous section with those of the \emph{Pseudoplatystoma} fishes. Finally, in Section \ref{sec:conclus}, we will present the final conclusions of this work.

%This is done trough extensive numerical simulations whose more important result are summarized in Section \ref{sec:numerical}. 
 %Furthermore, the influence of space dependent coefficients  an the inclusion of the advection  in the RD equations are explored. Finally, in Section \ref{sec:comparison}, we compare our patterns with those on the skin of \emph{Pseudoplatystoma} fishes and discuss the results.

%The purpose of this work is to delve into the BVAM model exposed in Section \ref{sec:bvam} in order to find the possible causes of the different patterns of the \emph{Pseudoplatystoma} species. 

%%%%%%%%%

%In general, the information that exists on the formation of those patterns and their diversity is very low. For example, it is known that the pigmentation begins to show at nine days of age. This appears, first, in the head part, until they are clearly defined in the body at the tenth day. See Figure \ref{fig:embrion}. It is also believed that the initial peculiar pattern of post-larvae is kept for two and a half months until it takes their final form. This pattern serves to camouflage the fish as a defense when they are dragged into the flooded river banks, where the vegetation provides ideal sites for concealment\cite{perez2001reproduccion}.

\section{The BVAM reaction diffusion model and its numerical approximation \label{sec:bvam}}

The interactions among morphogenes within living organisms  can be conceptualized as two coupled processes involving chemical reactions and differential diffusion. Thus, reaction diffusion models seem to be  appropriate quantitative schemes for reproducing the spatial distribution of these morphogenes previous to their manifestation as a given pattern on, for instance, the skin of organisms. In this context, a very useful model that allows one to obtain a wide variety of patterns is the BVAM model, proposed originally in \cite{barrio1999two}. In this model, the authors suggested that, near the equilibrium concentration of both morphogenes (located at the origin), the interaction between both components could be sufficiently described as a third order reaction. This reaction scheme, together with an unequal diffusion of both morphogenes, $u$ and $v$, gives place to a reaction-diffusion model which, in its dimensionless form, can be written as \cite{Barrio2009}:
\begin{subequations}\label{eq:bvam2}
\begin{equation}
\frac{\partial u}{ \partial t}=D \nabla^2 u+ \eta(u + \alpha v - \delta u v - u v^2),
\end{equation}
\begin{equation}
\frac{\partial v}{ \partial t}= \nabla^2 v+ \eta(\beta v + \gamma u + \delta u v + u v^2).
\end{equation}
\end{subequations}
%\begin{align}
%\\
%\end{align}

\noindent The spatial scale was normalized with the characteristic length of the domain whereas time was normalized with a combination of the diffusion coefficient of the morphogen $v$ and the spatial scale \cite{Barrio2009}. Thus,  $D=D_u /D_v$ is the ratio of the diffusion coefficients associated to both morphogenes,and $\eta$ provides the characteristic spatio-temporal scale of the pattern formation process. The  coefficients $\alpha$, $\beta$, $\delta$  and $\gamma$ are adjustable parameters which determine the appearance and form of the patterns. 

The idea of using a third order reaction scheme is that it provides three  different fixed points, providing of great richness to the variety of patterns obtainable by the model \cite{Barrio2009}. Besides the origin $(u_0,v_0)=(0,0)$, the additional two fixed points are  $(u_0,v_0)=(-\zeta,1)w_{\pm}$, where $w_{\pm}$ is given by:
\begin{equation}
w_{\pm}=\frac{-\delta\pm \sqrt{\delta^2+4\epsilon}}{2},
\end{equation}

\noindent and $\epsilon$ and $\zeta$ are defined as $\epsilon=\beta/\zeta-\gamma$ and  $\zeta=(\alpha+\beta)/(1+\gamma)$. 

%In order to perform an stability analysis of the system, one should consider the Jacobian matrix of this system:
%\begin{equation}
% \mathbb{J} = \left( \begin{array}{cc}
 %1-\phi & \alpha-\psi \\
% \gamma+\phi & \beta+\psi \\
% \end{array} \right),
%\end{equation}

%\noindent with $\phi=v_0(\delta+v_0)$ and $\psi=\delta u_0+ 2u_0 v_0=-\zeta v_0 (\delta+ 2 v_0)$, and also the  diffusion  matrix $\mathbb{D}$, which is a diagonal matrix defined by the diffusion coefficients $D_u$ and $D_v$. In this case, the linearizated system can be studied in term of the eigenvalues $\lambda_k$ of the matrix $\mathbb{J}-k^2\mathbb{D}$ for each individual fixed point. 

A detailed study of this model can be found in Ref.  \cite{varea2007soliton}. The main conclusion is that the model  produces different patterns, ranging from spots to strips, by just changing one single parameter, namely $\delta$, see Figure \ref{fig:barrio}. Furthermore, numerical simulations confirm that a great variety of stationary and transient patterns can be obtained from the BVAM scheme \cite{Barrio2009,barrio1999two}. Of particular interest to our work are 
the striped-like patterns, and those that contain some disorderly dots, as shown in Figure \ref{fig:barrio}, since they are present in \emph{Pseudoplatystoma} fishes.

%\subsection{Numerical validation of patterns}

In this study we have developed our own numerical solvers, employing different approximation schemes, mainly based on the finite difference method and on the finite element method \cite{thesis}. We have been able to reproduce several
patterns previously reported in the literature for the BVAM model and others, validating in this way our codes. Due to its better approximation properties and its flexibility to deal with boundary conditions, in the present study we employed a finite element method for space discretization, and a semi-implicit Euler scheme to integrate the discrete models in time.  This numerical scheme is implicit for the diffusion term and explicit for the reaction terms. The details can be found in \cite{thesis}.

\begin{figure}
\includegraphics[scale=0.18]{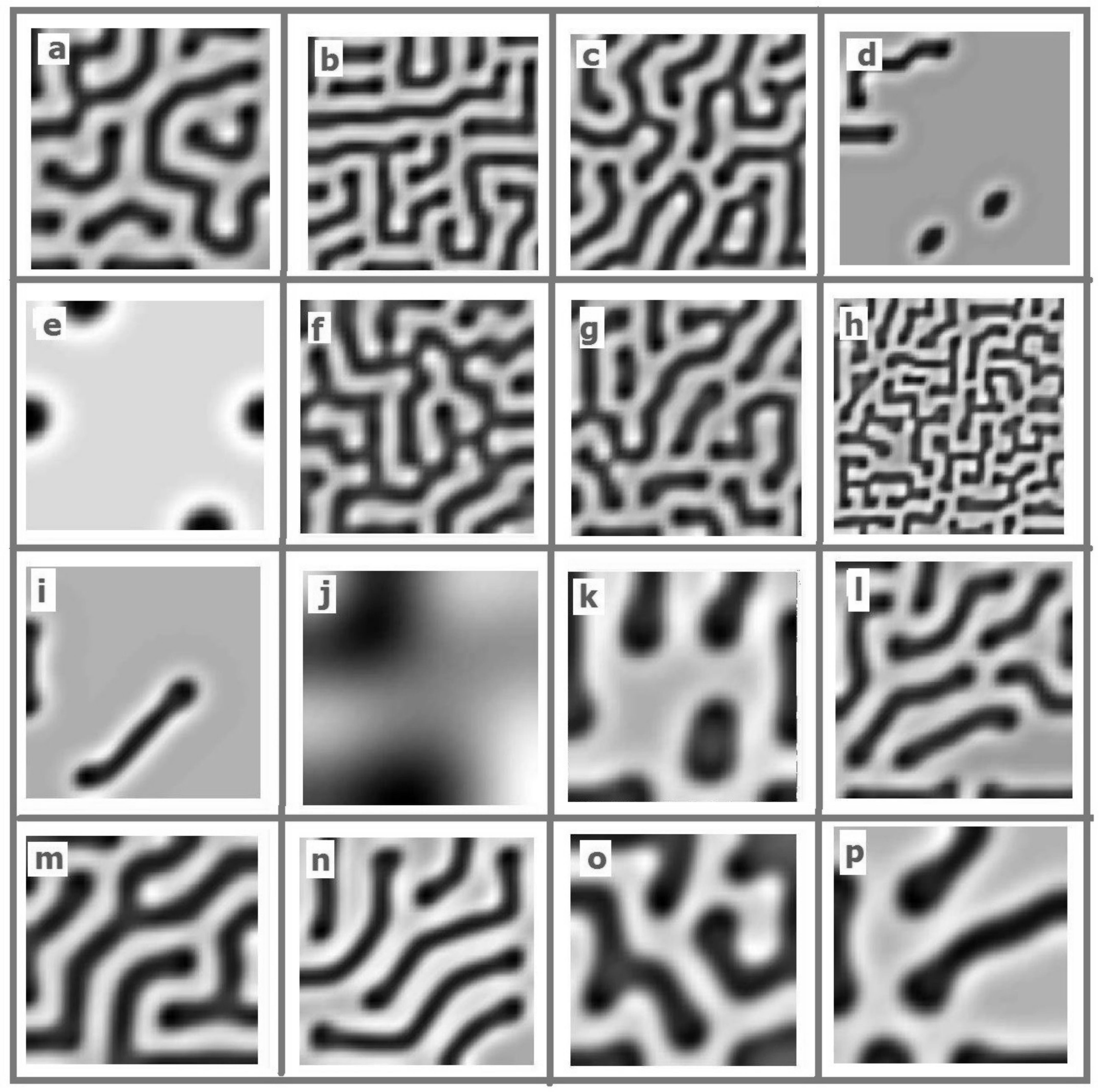}
\caption{Patterns (not necessarily stationary) of the BVAM model in a square domain of size $L=80$, after a time of T=750. Initial conditions are given by both concentrations near the equilibrium value $(0,0)$, with a random  addition of maximum amplitude $0.05$. Furthermore, we use zero flux boundary conditions at the four boundaries. The value of the parameters $\epsilon, \zeta$ and $\gamma$  are the same as the previous figure. The triad of parameters  $(D, \delta, \eta)$ are given as: a(0.344; 0.72; 0.225),	b(0.172; 2.0; 0.1),	c(0.172; 0.8; 0.15),	d(0.344; 1.5; 0.3), e(0.516; 0.72; 0.15),	f(0.172; 0.00; 0.15),	g(0.172; 0.72; 0.15),	h(0.172; 3.5; 0.15), i(0.344; 1.55; 0.15),	j(0.344; 1.7; 0.15),	k(0.344; 0.72; 0.075),	l(0.344; 0.0; 0.225), m(0.344; 0.0; 0.15),	n(0.344; 1.4; 0.15),	o(0.516; 0.0; 0.15),	p(0.516;0.36; 0.15). Numerical simulations were made using Finite Element method with a mesh of
 3200 triangles  and a step size of $\Delta t=5\times 10^{-2}$. The detail of the numerical method used in reaction-diffusion systems could be found in \cite{thesis}  . \label{fig:tapete}}
\end{figure}

In Figure \ref{fig:tapete}, we show a set of patterns extending those shown in Figure \ref{fig:barrio}, by considering a wider range of the parameters $D$, $\eta$, and $\delta$ in equations (\ref{eq:bvam2}). We have confirmed the appearance of dots, stripes and reticulated patterns for the BVAM model for conventional conditions, i.e., zero flux boundary conditions, random initial conditions near the equilibrium point (0,0), and constant kinetic parameters along the entire domain. 

However, it is convenient to emphasize that those computed patterns do not exhibit the preferential orientation  present in some of the \emph{Pseudoplatystoma} fishes. In the next section, we will focus our attention in the mechanisms inducing changes in a pattern obtained with some specific set of parameters. In particular, we will analyze the problem of how these changes could contribute to the orientation of the basic geometrical forms of a pattern.

\section{Patterns under non-conventional conditions \label{sec:numerical}} 

%\section{Numerical Results\label{sec:numerical}}

\subsection{Initial conditions}

It is noteworthy that, despite the great attention that reaction diffusion systems have had in recent years in the study of pattern formation, little has been studied about the influence of initial and boundary conditions  in the solution of the problem, although they are factors that complement the mathematical model and may play an important role in the type of solutions obtained \cite{arcuri1986pattern}. Our interest in this section is to examine how these two factors could produce a vertical orientation on the stripes, and to show how the variation of initial conditions could produce very different patterns, in spite of using the same parameter set.

The initial conditions  more often used in these types of problems are obtained by adding a random contribution to both concentrations placed at their equilibrium values (see for example \cite{barrio1999two,dufiet1992numerical}). In that case, the orientation of the striped-like patterns  can occur in at least two ways: 1) using boundary conditions of the Neumann type that aligns the stripes parallel or perpendicular to the boundaries, or 2) adding some preferential direction to the initial conditions \cite{dufiet1992numerical,madzvamuse2006time,ruuth1995implicit}. 

The influence of the initial conditions on the solutions obtained with the BVAM model of equations (\ref{eq:bvam2}) is shown in Figure \ref {fig:initial}. In this figure, the temporal evolution of the pattern is tracked from the initial condition (at the left of each row), until the stationary pattern (at the right), with three intermediate states. We emphasize that all other data in the model were maintained fixed, like the size of the domain, the boundary conditions (Neumann, zero flux) and the set of parameters. The specific information about the initial conditions used in each case is the following:

\begin{figure}
\includegraphics[scale=0.18]{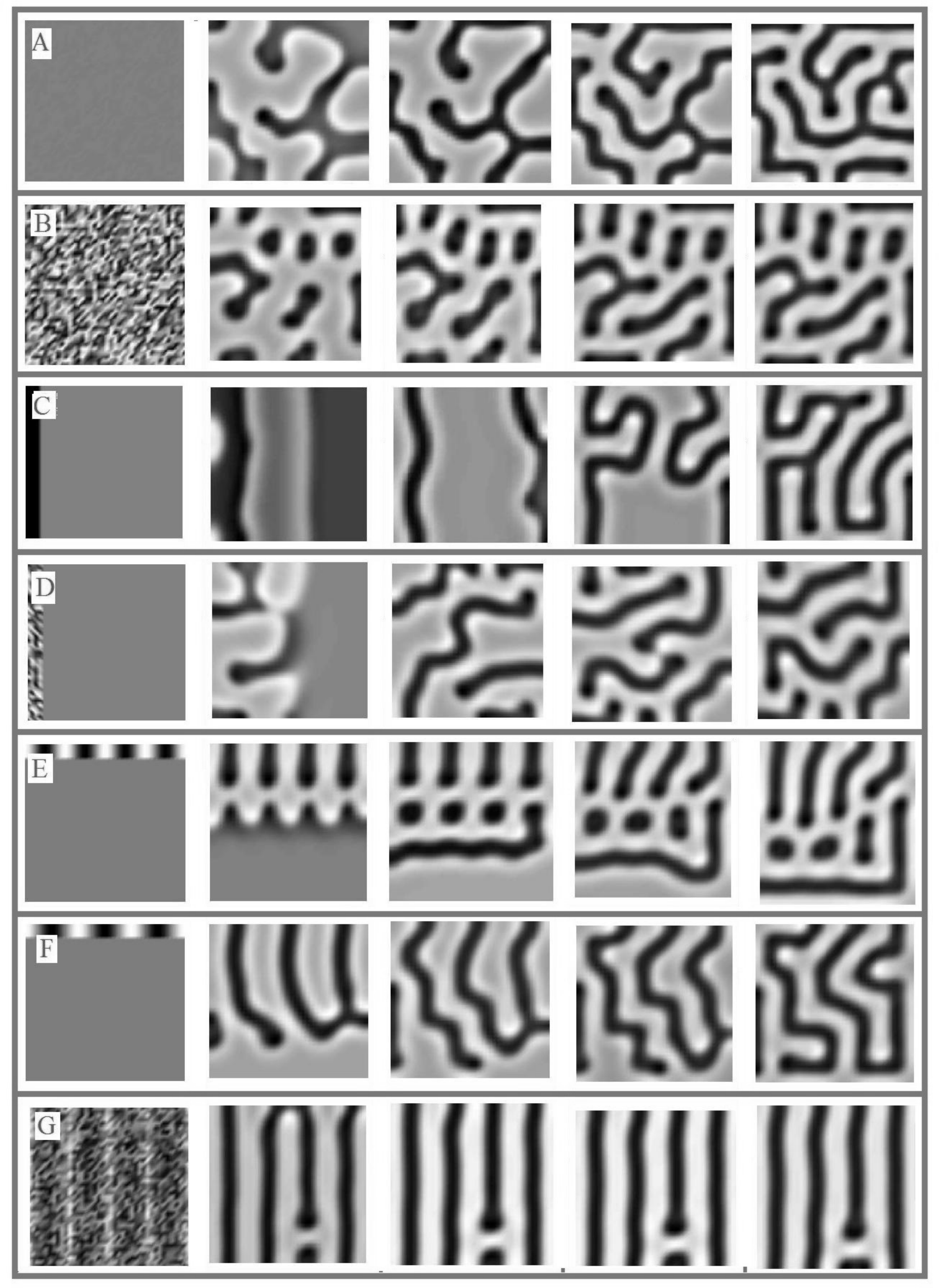}
\caption{Different patterns obtained varying only the initial conditions. The left column shows the initial condition; the intermediate columns show three intermediate states, and on the right the stationary pattern obtained after a time $T$, given in each case by: A) 875 B) 875 C) 5000 D) 875 E) 5000 , F) 4000, G) 500. The domain is also a square of side 80 and the reaction parameters were the same as Figure \ref{fig:barrio}.B for all cases. 
 \label{fig:initial}}
\end{figure}

\begin{enumerate}[leftmargin=0cm,itemindent=1.3cm,
  labelindent=1.0cm,labelwidth=\dimexpr1cm-.5em\relax,
  labelsep=!,align=left, label=\roman*]
  \item.- We first compare the rows \ref{fig:initial}.A and \ref {fig:initial}.B where we have used conventional random initial conditions, with the only difference that the noise intensity is different in each case. In Figure \ref{fig:initial}.A, the initial values $u$ and $v$ are between $\pm 0.001$, whereas in \ref{fig:initial}.B are between $ \pm 0.1$. The results show that the obtained stripes  are (on average) longer in the former case than in the latter, where some spots look more like dots than as stripes.
  \item.- Now, we compare the rows \ref{fig:initial}.C and \ref{fig:initial}.D. In both cases, the concentrations of the two morphogenes are zero except in a small strip on the left side of the square. In the first case, the initial condition is a left vertical stripe with 
$u = -0.1 $ and $v$ randomly distributed, while in the second case, both concentrations are random with a maximum amplitude of $0.05$. In both cases, the disturbance spreads to the other side of the domain. However, the wave front in \ref{fig:initial}.C proceeds as a vertical strip, which is duplicated and bounces off at the right vertical boundary to form a labyrinth type form. In the second case, \ref{fig:initial}.D,  the disturbance advances in disorder, even with a slight preference for the horizontal direction, giving, at the end, a similar pattern to \ref{fig:initial}.B, but more disorganized.
  \item .- Next, we compare Figures \ref{fig:initial}.E and \ref{fig:initial}.F. In this case, the initial concentration of both morphogenes is zero almost everywhere, except for a stripe at the top of the domain. In the first case, this strip consists of four cosine oscillations of $ u $ while the other has only three. The amplitude of the oscillations is $0.05 $. The variable $v$ is randomly distributed in this strip and zero outside. In the case of \ref{fig:initial}.E, stripes and spots move downward preserving the same arrangement of columns, whereas in the case \ref{fig:initial}.F  the strips disperse and form a pattern similar to that in \ref{fig:initial}.C.  Other choices of the number of initial oscillations leads to results similar to those of \ref{fig:initial}.F. 
  \item .- Finally, we present an initial pattern with four stripes completely ordered, as shown in Figure \ref{fig:initial}.G. These initial conditions consist of four  vertical oscillations, plus random noise for both variables. These oscillations are taken with a maximum amplitude of 0.05. The almost perfect vertical orientation vanishes when the number of initial oscillations is different to four. In this case, the stripes fade, forming a pattern rather similar to that in Figure \ref{fig:initial}.D, without any preferential orientation. 
\end{enumerate}

The previous results allow us to conclude that: 1) Neumann boundary conditions tend to align the patterns with the edges; 2) the forms of the pattern  may travel as a perturbation and stay or fade depending upon the initial condition; and finally, 3) the results obtained in \ref{fig:initial}.E and \ref{fig:initial}.G suggest that only specific initial conditions (an ordered chemical pre-pattern) of the morphogenes can yield specific orientations of the stripes, like those observed in live organisms. This suggests, in turn, that a previous chemical reaction diffusion mechanism may be necessary to establish an ordered chemical pattern. In this form, the development of a complex organ, like the skin, occurs as a series of stages that involve different time scales and activation/deactivation subprocesses. 

\subsection{Boundary conditions}

Another aspect to consider is the type of boundary conditions used. Conventionally, either zero flow or periodic boundary conditions have been imposed on all the boundaries of the domain  to demonstrate the existence of patterns \cite{dufiet1992numerical,madzvamuse2006time}. In our case, considering the form of \emph{Pseudoplatystoma} fish, we may assume that a combination of both types of boundary conditions could model better the spatial disposition of the resulting patterns along  the body of the fish. Because the pattern in the fishes are spatially repeated from the end of the head (operculum) to the root of the tail (caudal fin), see Figure \ref{fig:surubies}, we propose a domain with no flow through the horizontal boundaries (top and bottom), and periodic conditions on the vertical edges.

\begin{figure}
\includegraphics[scale=0.30]{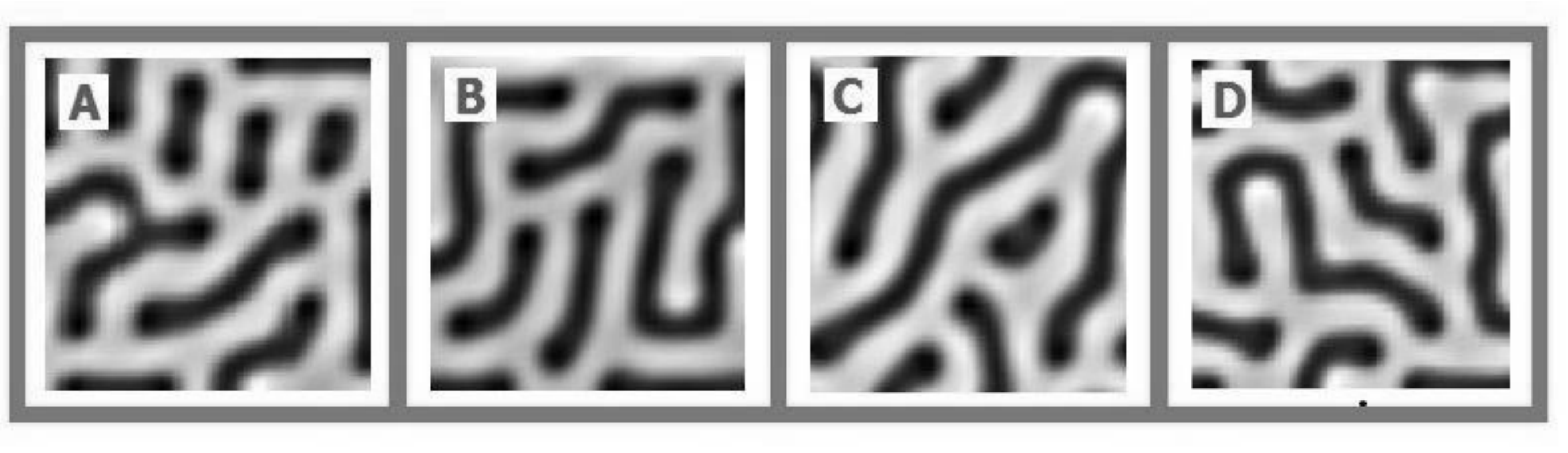}
\caption{Patterns obtained by varying only the boundary conditions. In each case, these are given by: A) Neumann, zero flux, B) Neumann, with a current source 0.05 in the upper border, C) periodic, and D) Left and right edges are periodic, and zero flux up and down. Patterns are taken after a time T = 350. The parameters are the same as in Figure \ref{fig:barrio}.B. \label{fig:frontier}}
\end{figure}

However, as we show in Figure \ref{fig:frontier}, the election of the boundary conditions not necessarily affects the type of pattern obtained. For the results shown in this figure, we have employed four combinations of boundary conditions without any significant alteration of the geometrical form (stripes) obtained in Figure \ref{fig:barrio}.B. The same statement is also true for the other three kind of patterns of the Figure \ref{fig:barrio}.  Therefore, we conclude that the election of boundary conditions does not significantly affect the formation and appearance of the pattern, at least for the cases we have studied here.

\subsection {Space dependent coefficients}

In this section, we will explore the spatial variation of the parameters of the chemical reaction of the BVAM model. The aim is to create combinations of the basic patterns of the BVAM model,  which are sown in Figure \ref{fig:barrio}, in order to obtain the complex patterns appearing in the \emph{Pseudoplatystoma} fishes shown in Fig. \ref{fig:surubies}. 

To achieve this objective, we have varied the coefficients  $\eta$ and $\delta$ spatially. We select these parameters because it is well known how the shape of the patterns changes with their variation. We want to make clear that we will not consider a spatial variation of the diffusion coefficient $D$. Nevertheless, in previous works it has been proven that an anisotropic diffusion can generate orientated patterns \cite{sanderson2006advanced,shoji2002directionality}. Those models assume that each of the substances have a preferential direction to spread.

To construct a spatial variation of $\delta$, we consider the four different patterns that arise with the BVAM model when different constant values of this parameter are imposed \cite{Barrio2009}. Those patterns are 1) reticulated, 2) stripes, 3) dots and stripes, and 4) disordered dots, and they are obtained at specific values that we have labeled as $\delta_1$, $\delta_2$, $\delta_3$, $\delta_4$, respectively, as shown in Figure \ref{fig:barrio}. The idea in this section is that the transition between a specific kind of pattern and other in the same figure could be obtained by changing the value of the parameter $\delta$ spatially.

%As Barrio noted, all the basic forms may be obtained by varying only the quadratic coefficient $\delta$ (See Figure \ref{fig:barrio}). Therefore we will explore how to make a transition between  different kind of forms in order to reproduce the variability of shapes in the skin of the \emph{Pseudoplatystoma}. 

On the other hand, to construct a spacial  variation of $\eta$, we have to take into account that, in the dimensionless form of a reaction diffusion system, like that of equations (\ref{eq:bvam2}), the value $\eta$ provides a spatio temporal scale proportional to $L^2/t $, in such a way that, if the size of the domain $L$ is fixed,  increasing  $\eta$ represents a decrease in the size of the pattern shape and an increase in the reaction rate \cite{Murray2003}.

The results of our numerical experiments, when spatial variation is incorporated to $\eta$ and $\delta$, are summarized in Figure \ref{fig:espacio}, which we proceed to explain row by row:

\begin{figure}
\includegraphics[scale=0.3]{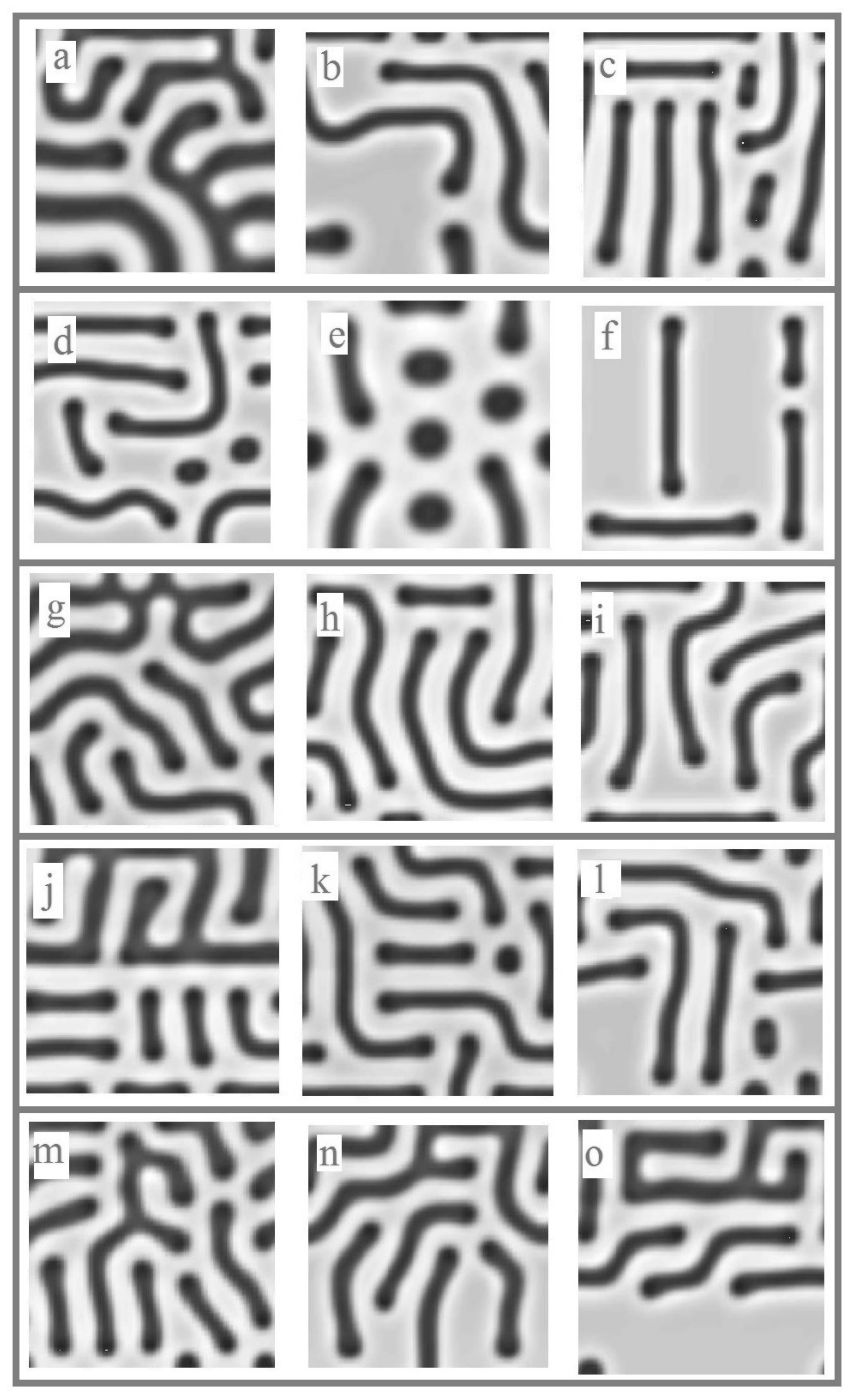}
\caption{Patterns with spatial dependence in the parameters $\eta$ and $\delta$. The other parameters are the same as Figure \ref{fig:barrio}. See text for details. \label{fig:espacio}}
\end{figure}

\begin{enumerate}[leftmargin=0cm,itemindent=1.3cm,
  labelindent=1.0cm,labelwidth=\dimexpr1cm-.5em\relax,
  labelsep=!,align=left,label=\roman*]

  \item.- The first row of Figure \ref{fig:espacio} presents the results obtained when $\eta$ increases linearly according to 
  
  \begin{equation}\label{eq:4}
\eta(y)=0.15+(0.40-0.15)\frac{y}{L}.
\end{equation}  
  
  The other parameters are the same as those used to obtain the results shown in Figures \ref{fig:barrio}.A, \ref{fig:barrio}.B and \ref{fig:barrio}.D, respectively, where $\eta$ had the fixed value of 0.35. It is clear that an increase of $\eta$ causes a change in the size of the spots (Figure \ref{fig:espacio}.a) and stripes (Figure \ref{fig:espacio}.b) without a significant change in form. However, in the case of dots, the gradient of $\eta$ causes that they become stripes (Figure \ref{fig:espacio}.c).

% we explore the effect of vary the parameter $\eta$ in the patterns of Figure \ref{fig:barrio}. In Figure \ref{fig:barrio} the value of $\eta$ was fixed at 0.35. In Figures \ref{fig:espacio}.a, \ref{fig:espacio}.b, and \ref{fig:espacio}.c, we use the same parameters  corresponding to Figures \ref{fig:barrio}.A, \ref{fig:barrio}.B and \ref{fig:barrio}.D respectively, (i.e. corresponding to $\delta_1$, $\delta_2$ and $\delta_4$), except that $\eta$ is no longer constant, but increase linearly from bottom to top in the form:
%  

  \item.- In the second row of Figure \ref{fig:espacio}, we show results obtained when  parameter $\delta$ oscillates around two typical values for which stripes and dots are obtained with the BVAM model, (i.e., around $\delta_3=0.71$ and $\delta_5=0.80$, respectively).  These oscillations are imposed  in the horizontal direction in the form:
  
%\begin{equation}
%\delta(x)=\delta_3+\frac{1}{2}(\delta_5-\delta_3) \left[1+\cos{\left(\frac{2\pi m x}{L} \right)} \right],
%\end{equation}  

\begin{equation}
\delta(x)=\delta_3+(\delta_5-\delta_3) \sin^2{\left(\frac{\pi m x}{L} \right)} ,
\end{equation}

\noindent where $m=2,3,4$ determines the number of oscillations of $\delta$ for Figures  \ref{fig:espacio}.d,  \ref{fig:espacio}.e, and  \ref{fig:espacio}.f, respectively. It is noted that the vertical orientation is retained only for the cases \ref{fig:espacio}.e and   \ref{fig:espacio}.f, while in Figure \ref{fig:espacio}.d, the spots and stripes are intermingled haphazardly.

   \item.- In the third row of Figure \ref{fig:espacio}, we present results that show the effect of decreasing $\delta$ linearly  between two different typical values where a pattern change occurs, see Figure \ref{fig:barrio}. A spatial variation of $\delta$ in the vertical direction is given by

\begin{equation}
\delta(y)=\delta_{n+1}+(\delta_{n} -\delta_{n+1})\frac{y}{L}, 
\end{equation}   
  
\noindent with $n=1,2,3$ for Figures \ref{fig:espacio}.g, \ref{fig:espacio}.h, and \ref{fig:espacio}.i respectively. We expected to obtain the following transitions: from spots to stripes, from stripes to stripes and dots, and finally, the transition from stripes to just dots, respectively. However, this did not occur due to the fact that the gradient of $\delta(y)$ tends to elongate all the forms into stripes.

   \item.- The fourth row is the same as above, except that now the variation of $\delta$ in the vertical direction is given by a jump at $y=L/2$. More specifically:
   
\begin{equation}
\delta(y) = \begin{cases} 
\delta_n & \mbox {if } y \geq L / 2, \\
\delta_ {n + 1} & \mbox{if } y <L / 2. 
\end{cases}    
\end{equation} 

\noindent  with $n=1,2,3$ for Figures \ref{fig:espacio}.j, \ref{fig:espacio}.k, and \ref{fig:espacio}.l respectively. This particular form of the parameter $\delta$ causes the overlapping of shapes and the distortion of the dots, except in Figure \ref{fig:espacio}.j where there are two distinct regions. This is because the difference between $\delta_1$ and $\delta_2$ is much grater (in absolute value) than in the other cases.

      \item.- Finally, in the last row of Figure \ref{fig:espacio}, we present combinations of the previous cases. 
      
      In Figure  \ref{fig:espacio}.m, we present the result obtained with a linear change of $\delta$, which includes the values covering the entire range from spots to dots:
\begin{equation}
\delta(y)=\delta_4+(\delta_1-\delta_4)\frac{y}{L}.
\end{equation}

In Figure \ref{fig:espacio}.n we present the result obtained with a variation of $\delta$, combining oscillations in the horizontal direction (like those of Fig. \ref{fig:espacio}.e), and a linear decrease in the vertical direction (like that of  Fig. \ref{fig:espacio}.g). For this last case, the spatial variation of $\delta$ is given by

\begin{equation}
\delta(x,y) =\left[(\delta_4 -\delta_1) - (\delta_4- \delta_2)\cos^2{\left(\frac{3\pi x}{L} \right)}\right]\frac{y}{L}.
\end{equation}

Figure \ref{fig:espacio}.o is similar to Figure \ref{fig:espacio}.m but now the spatial variation of $\delta$ in the vertical direction is a piecewise constant function defined by:

\begin{equation}
\delta(y) = \begin{cases} \delta_1 &\mbox{if } y\geq 2L/3, \\ 
\delta_2 & \mbox{if } 2L/3>y>L/3. \\
\delta_4 & \mbox{if } y\leq L/3.\end{cases}.
\end{equation}

In these last three cases, we obtained the combination of different forms in the same pattern, and is noteworthy that the change is more marked when the coefficients are discontinuous.

\end{enumerate}

From the results of this part of the section, we can conclude that the spatial variation of the kinetic parameters allows us to obtain a somewhat larger combination of patterns. However, the occurrence of stripes with dots in the same pattern is very difficult to obtain by following this strategy. One of the reasons is that the gradient of one parameter  ($\delta$ or $\eta$ in our case) tends to elongate the dots and turn them into stripes. Therefore, in the next section, we will consider another possibility for obtaining the combination of stripes and dots, and a preferential orientation in a pattern. 

\subsection{Effect of advection}

\begin{figure}  
\includegraphics[scale=0.30]{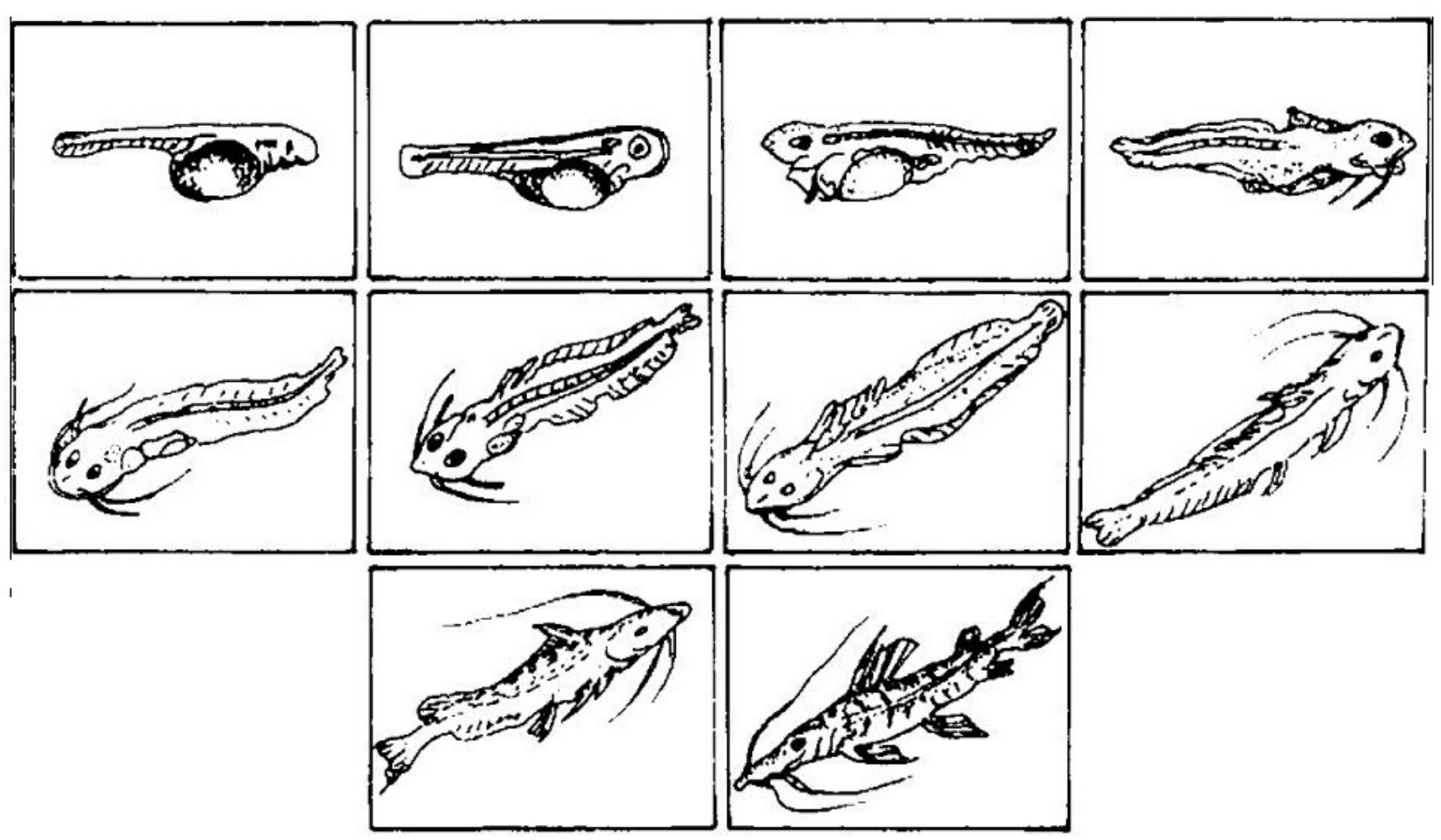}
\caption{Stages of larval development. It is up to the ninth day that the pigmentation start to be noticed, mainly in the head. Taken from Ref. \cite{perez2001reproduccion} \label{fig:embrion} }
\end{figure}

In general, the information that exists about the mechanisms that lead to the formation of the patterns of the \emph{Pseudoplatystoma} and their diversity is very scarce. For example, it is known that the pigmentation begins to show at nine days of age. This pigmentation appears, first, in the head part, until it is clearly defined in the body at the tenth day, see Figure \ref{fig:embrion}. It is also believed that the initial peculiar pattern of post-larvae is kept for two and a half months until it takes its final form. This pattern serves to camouflage the fish as a defense when they are dragged into the flooded river banks, where the vegetation provides ideal sites for concealment \cite{perez2001reproduccion}.

The fact that the coloration appears first in the head, and then is dispersed along the entire skin, suggests the possibility of an underlying transport process besides the  reaction diffusion interaction. In this section, we explore the possibility that the interaction between both morphogenes may be influenced by an advective sub-process along the fish skin. Specifically, we will show that the advection could be seen as a possible origin of the preferential orientation of some fish patterns.

In view of these considerations, we add an advective term to the original reaction diffusion system described  in equations (\ref{eq:bvam2}), obtaining as a result a reaction convection system for the BVAM model of the form \cite{galeano2013formacion}:

\begin{subequations}\label{eq:convect}
\begin{equation}
\frac{\partial u}{ \partial t}+ \mathbf{z}\cdot \nabla u=D \nabla^2 u+ \eta(u + \alpha v - \delta u v - u v^2),
\end{equation}
\begin{equation}
\frac{\partial v}{ \partial t}+ \mathbf{z}\cdot \nabla v= \nabla^2 v+ \eta(\beta v + \gamma u + \delta u v + u v^2).
\end{equation}
\end{subequations}

\noindent We are particularly interested in a vertical advective flux through the term $\mathbf{z}=z\hat{e_y}$, with $z$ a constant parameter for the volumetric velocity of both morphogenes, and $\hat{e}_y$ the unitary vector in the vertical direction.

\begin{figure}
\includegraphics[scale=0.22]{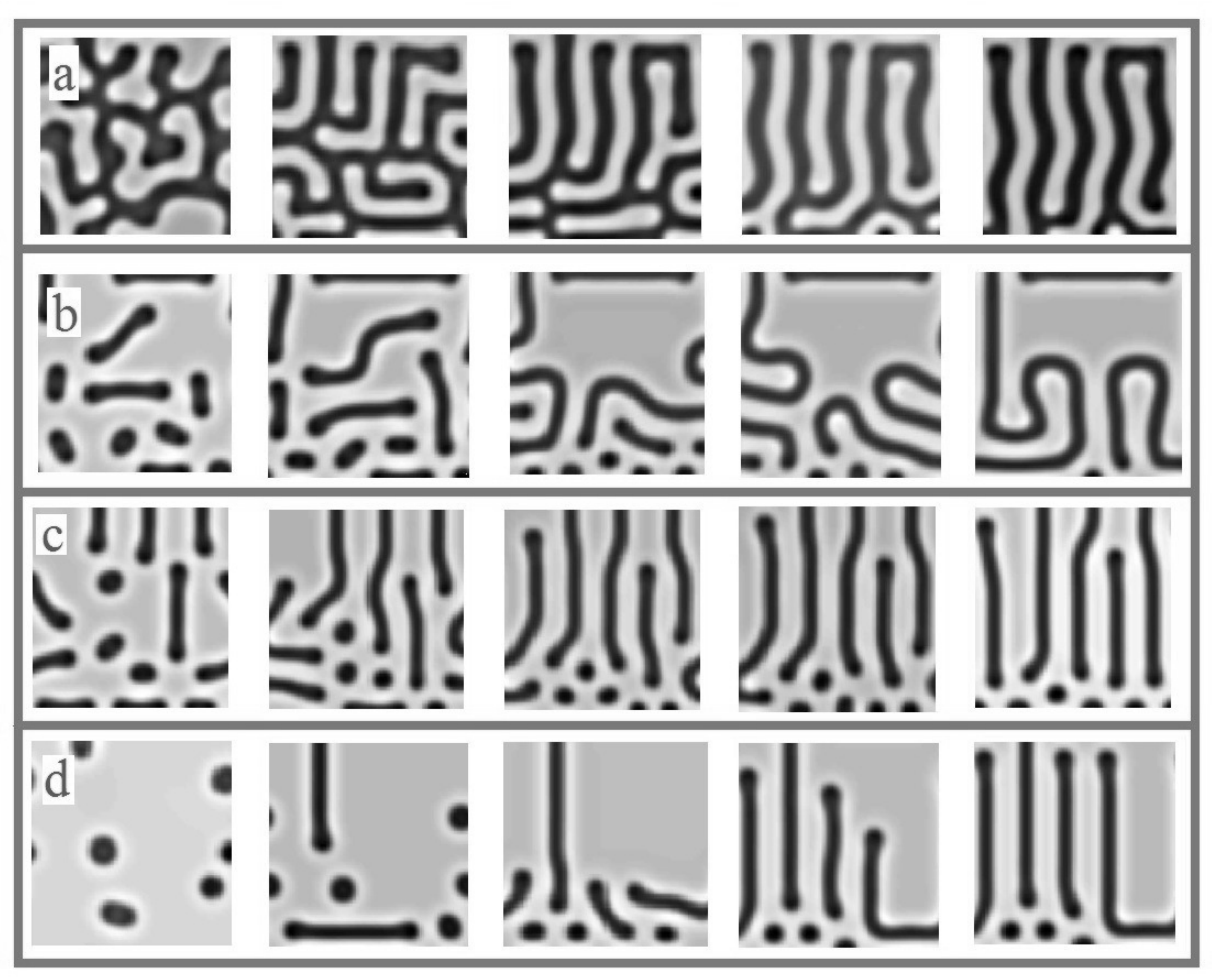}
\caption{Results obtained with the same four set of parameters of Figure \ref{fig:barrio}, but with the effect of advection in the vertical direction. For these simulations, we used $\mathbf{z}=0.02\hat{e}_y$ in equations (\ref{eq:convect}). At the beginning (left column), the pattern resembles the reaction diffusion pattern of Fig. \ref{fig:barrio}, but as time goes, the advection aligns the forms (right column). The simulations use zero flux boundaries conditions at the top and bottom boundaries, and periodic boundary conditions at both vertical sides. \label{fig:adveccion}}
\end{figure}

In Figure \ref{fig:adveccion}, we show the results obtained by adding the effect of advection to the original patterns of Figure \ref{fig:barrio}. As it is shown,  advection in the vertical direction introduces not only the desired  orientation on the patterns, but also the combinations of dots and stripes, as we have previously searched for some of the \emph{Pseudoplatystoma} fishes.

We can see that, for all cases, the orientation of the pattern tends to align along the direction in which advection takes place. Also, in the same figure, it can be observed that dots are present in the final pattern, at the end of the lines, as long as somehow they are present in the parameter set chosen.

\section{Comparison with patterns in \emph{Pseudoplatystoma} fishes\label{sec:comparison}}

In this section we compare the lateral region of the \emph{Pseudoplatystoma} fishes with some of the  most similar patterns obtained previously with numerical solutions of the BVAM model, see Figure \ref{fig:minipaterns}:

\begin{figure}
\includegraphics[scale=0.36]{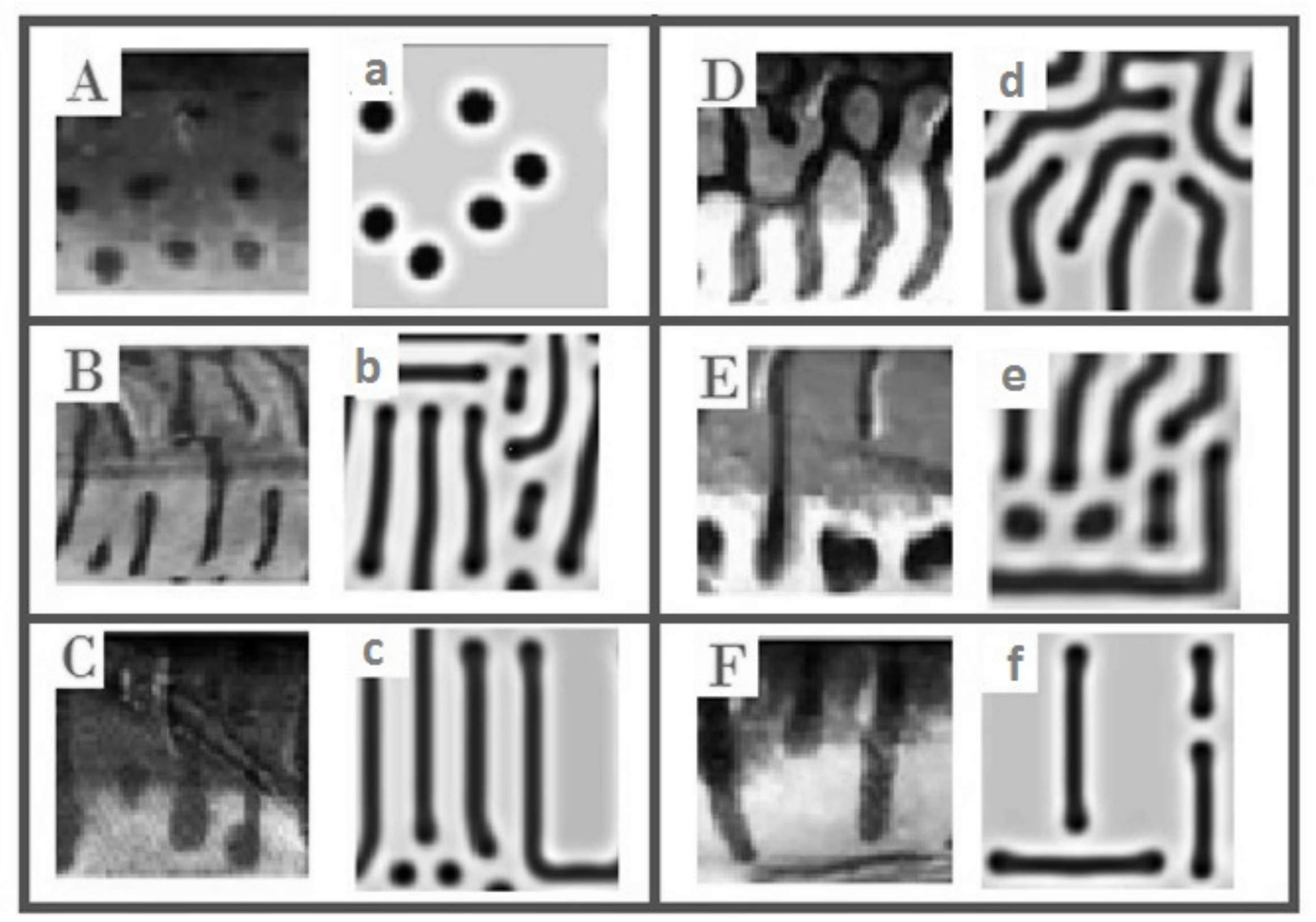}
\caption{Comparison between the pattern in the skin of the \emph{Pseudoplatystoma fishes} (capital letters) and some of the patterns derived  from numerical simulation of the BVAM model (lower case letter). See text for details.  \label{fig:minipaterns}}
\end{figure}

\begin{enumerate}[leftmargin=0cm,itemindent=1.3cm,
  labelindent=1.0cm,labelwidth=\dimexpr1cm-.5em\relax,
  labelsep=!,align=left,label=\roman*]
  \item.- In the case of \emph{P. coruscans} (Figure \ref{fig:minipaterns}.A), the pattern presents semicircular spots horizontally staggered, with some elongated spots on the top. This suggests that these patterns can be obtained with the conditions of Figure \ref{fig:barrio}.D, but with some kind of horizontal dependency in the parameters to give regularity in the  horizontal direction.
  
  \item.- In the case of \emph {P.  reticulatum} (Figure \ref{fig:minipaterns}.B), the pattern is a reticulated on top, switching to vertical lines  at the bottom. These patterns could be reproduced with a vertical dependence in parameter $\delta$ as that of Figure \ref{fig:espacio}.n, or a combination 
of a spot pattern with advection as in Figure \ref{fig:adveccion}.a.

  \item.- In the case of \emph{P. tigrinum} (Figure \ref{fig:minipaterns}.C), the pattern includes interlaced vertical lines and dots. This pattern could be formed by initial dots and stripes in the presence of advection as those of Figures \ref{fig:adveccion}.c or \ref{fig:adveccion}.d

  \item.- In cases of \emph{P. oriconoense} in  Figures \ref{fig:minipaterns}.D, the pattern consist of reticulated spots at the top with vertical stripes at the bottom. Here, a vertical dependence in the parameter $\delta$ as in Figure \ref{fig:espacio}.n could be used.

\item.- In the case of \emph{P. fasciatum} in Figure \ref{fig:minipaterns}.E, the patterns consist of stripes and large dots at the bottom. These patterns can be obtained from a stripe pattern with special initial conditions, as the one in Figure \ref{fig:initial}.E.

  \item.- In the case of \emph{P. magdalenatium} in Figure \ref{fig:minipaterns}.F, the pattern consists of horizontal and vertical lines. We have achieved  this particular configuration with a horizontal dependence of $\delta$ as in Figure \ref{fig:espacio}.f.
 
\end{enumerate}

The previous results  allows us to conclude that the origin of the different patterns observed in fishes cannot necessarily be reproduced only by changing the parameters in the reaction diffusion scheme. In other words, the genetic aspects associated to the differentiation of patterns are not totally summarized in the specific value of the kinetic parameters of the reaction-diffusion system. 

This study permit us to suggest two hypothesis:
1) some morphological aspects that affect the size, the type and the condition of skin development, for different species, could influence its pattern formation; those aspects, whichever they are, not necessarily are implicitly contained in the reaction-diffusion scheme but can be incorporated through the initial and boundary conditions, as well as with the advective effects and the spatial dependence of the parameters. 2) Different patterns between individuals of the same species could be related to minor skin differences, associated to its genesis and development; these aspects depend on the particularity of each individual.

It is important to mention that, besides the factors considered in this work, we believe that considering a growing domain may also be a determining factor in the form and specificity of the pattern of each animal. This because  the formation of patterns in the skin occurs at the same time that its growth, and it is well known that the resulting pattern in a reaction diffusion system depends upon the size of the domain \cite{crampin1999reaction,madzvamuse2005moving,painter1999stripe}

\section{Conclusions\label{sec:conclus}}

In this work, we have shown that reaction diffusion models for the interaction of morphogenes may reproduce more readily  and accurately the skin patterns observed in animals (such as the \emph{Pseudoplatystoma} catfishes) when some factors, usually not included in the reaction-diffusion equations, are considered. We have showed that boundary and initial conditions, spatial dependency of the parameters and the effect of the advection, could be very important factors in the determination of the mechanism of pattern formation. 

This work compel us to investigate not just the set of parameters that reproduce some specific pattern, but also other possible sub-processes present in the morphogenesis, for instance, factors that affect the biophysical domain where the process takes place \cite{madzvamuse2006time,madzvamuse2005moving}. More important seem to be the specific conditions at which initiates the process of interaction between those morphogenes and its surroundings \cite{Kondo1995}.

Given the variety of models which have been introduced to investigate pattern formation, this work emphasizes that the choice of one model should be based not only on their ability to reproduce a specific pattern, but also the evolution of the pattern throughout the differentiation process. Also, the election may take into account the ability of the model to reproduce the differences between organisms of the same species, study that requires a quantitative analysis of sensitivity to initial parameters and geometric conditions in a bidimensional  system \cite{arcuri1986pattern}.

However, the achievement of this goal requires not only the help of computer simulations, but more importantly, it depends upon obtaining enough biological data that allow us, firstly, to track the morphogenes involved in the differentiation process, and then, the spatio-temporal evolution of the patterns on a specific organism. Without this experimental information, direct comparison of the biological patterns with those obtained with a specific reaction diffusion system, cannot be justified on terms other than mere coincidence.

\begin{acknowledgements}
A.L.D. acknowledges CONACyT for financial support under fellowship 221505 and the support provided by the graduate program of the Department of Mathematics at UAM-Iztapalapa. I.S.H. acknowledges financial support from DGAPA UNAM under grant IN113415.
\end{acknowledgements}

%%%%%%%%%%%%%%%%%%%%%%%%%%%%%%%%%%%

% BibTeX users please use one of
%\bibliographystyle{spbasic}      % basic style, author-year citations
\bibliographystyle{spmpsci}      % mathematics and physical sciences
\bibliography{ref4.bib}   % name your BibTeX data base

%% Non-BibTeX users please use
%\begin{thebibliography}{}
%%
%% and use \bibitem to create references. Consult the Instructions
%% for authors for reference list style.
%%
%\bibitem{RefJ}
%% Format for Journal Reference
%Author, Article title, Journal, Volume, page numbers (year)
%% Format for books
%\bibitem{RefB}
%Author, Book title, page numbers. Publisher, place (year)
%% etc
%\end{thebibliography}

\end{document}